\documentclass[conference]{IEEEtran}
\ifCLASSINFOpdf
\else
\fi
\usepackage{amsmath}
\usepackage{multicol}
\usepackage{lipsum}
\usepackage{graphicx}
\usepackage{subcaption}
\usepackage{cite}
\usepackage{url}

\begin{document}
%
\title{The network footprint of replication in popular DBMSs}


%
\author{\IEEEauthorblockN{Muhammad Karam Shehzad,
Jam Muhammad Yousif,
Muhammad Saqib Ilyas, and
Adnan Iqbal}
\IEEEauthorblockA{Namal Institute, Mianwali, Pakistan\\ \{karam.shehzad@hotmail.com,\,jammuhammadyusif@gmail.com,\,msaqib@gmail.com,\,adnan.iqbal@namal.edu.pk\}}}


\maketitle

\begin{abstract}
Database replication is an important component of reliable, disaster tolerant and highly available distributed systems. However, data replication also causes communication and processing overhead. Quantification of these overheads is crucial in choosing a suitable DBMS form several available options and capacity planning.
In this paper, we present results from a comparative empirical analysis of replication activities of three commonly used DBMSs - MySQL, PostgreSQL and Cassandra under text as well as image traffic. In our experiments, the total traffic with two replicas (which is the norm) was as much as $300$\% higher than the total traffic with no replica. Furthermore, activation of the compression option for replication traffic, built in to MySQL, reduced the total network traffic by as much as $20$\%. We also found that average CPU utilization and memory utilization were not impacted by the number of replicas or the dataset.
\end{abstract}

%
\IEEEpeerreviewmaketitle
\section{Introduction}
\label{sec:intro}
Most of today's distributed systems and applications have redundancy in hardware and software. This hardware/software replication helps provide high availability, i.e., service continuation in the face of component failure. Many service providers place their replicas at geographically disparate locations so that they are less likely to be affected by a natural or man-made disaster. If one site is taken out by a disaster, another replica can continue to serve client requests. This is known as disaster recovery. Geographic dispersion of replicas also lowers an average user's latency to the service as a replica is expected to be available nearby. In short, replication assists delivery of highly available, disaster tolerant and low latency services. 

The benefits of replication must be balanced against its costs. In the context of disaster recovery, Wang et al. described replication in terms of three phases, namely, deployment, synchronization and failover~\cite{wang:2015:dsn}, each of which has costs associated with it. In the deployment phase, the capital cost of hardware~\cite{greenberg:2008:CCR} is unavoidable. The software replication setup may also be performed at system deployment time by copying initial application and data over at all replicas. However, in the context of cloud computing, it is increasingly being done by live migration of virtual machines. Different authors have suggested a minimum bandwidth requirement of between $90$ Mbps and $680$ Mbps~\cite{VMWareCisco:2009,VMWarevSphere:2015,mashtizadeh:2014:xvmotion,oberg:2011:CommMag,cully:2008:usenix,wang:2015:dsn,kokkinos:2016:CSUR} to support live migration of VMs. During the synchronization phase, the replicas are kept consistent. The operational costs of replication during synchronization include real estate leasing costs, electricity costs and employee salaries etc~\cite{greenberg:2008:CCR}. The fail-over phase kicks in when the primary site fails and the replica or secondary site must take over serving client traffic. While the secondary site takes over, some client traffic might be lost, resulting in revenue losses. 

In this paper, we focus on the network traffic on expensive inter-data center links during the synchronization phase. This traffic is necessary to keep the replicas consistent with the primary or master server. Since network transit costs are known to be a significant fraction of data center operations cost~\cite{greenberg:2008:CCR}, this traffic must be kept to a minimum. At the very least, knowledge of the expected traffic volume can be useful in planning and provisioning of inter-data center links.

A significant contribution to the replication traffic is expected from updates to application databases due to user interaction with hosted applications, such as uploading photos or posting status updates.  Most commonly used Database Management Systems (DBMSs) provide support for replication setup and synchronization. However, the mechanisms and overhead associated with replication differ among these systems. 

In this paper, our primary focus is to evaluate three popular DBMSs, namely, MySQL, PostgreSQL and Cassandra, in terms of their network level footprint of replication for different types of update traffic. However, if a particular DBMS has a lower network utilization for replication, we are also interested to know if it trades off network utilization over some other metric, such as CPU or memory utilization. To this end, we use a variety of real traces of updates to a database master server, while capturing the traffic generated from the master server to the replicas. 

We found that the network traffic grew almost linearly with increase in number of replicas. For MySQL, an option to compress replication traffic, at the expense of slightly increased average CPU utilization, was available out of the box. This option was not available for the versions of PostgreSQL and Cassandra that were available at the time of our study. Enabling compression of replication traffic for MySQL reduced the total network traffic by about $20$\% for the textual data-sets and about $3$\% for the images data-set. 
We found that Cassandra had the least traffic overhead for the image data-set. It also had the best record insertion processing rate for all types of traffic. So, for image intensive applications or if sheer processing rate is what matters most, then Cassandra may be the DBMS of choice. However, PostgreSQL may have near optimal performance in terms of traffic overhead and record insertion rate. PostgreSQL not only has traffic encryption by default but also showed the lowest average CPU utilization during our experiments resulting in a lower energy consumption. Thus, if performance and energy consumption are both important, then PostgreSQL may be the DBMS of choice.

The rest of the paper is structured as follows. In section~\ref{sec:related}, we discuss related work, before describing the experimental setup of our study in section~\ref{sec:setup}. We present the results of our study in section~\ref{sec:results}. We make some recommendations in section~\ref{sec:discussion} and draw conclusions in section~\ref{sec:conclusions}.
\section{Related Work}
\label{sec:related}
Wiesmann et al. presented a classification and comparison of various database replication techniques in~\cite{wiesmann:2000:SRDS}. They used three criteria for comparison of replication techniques. The first is server architecture which dictates whether updates are applied to the primary only, at first, or to all replicas simultaneously. The second criterion is server interaction which dictates how frequently the servers communicate with each other. The third parameter is transaction termination which dictates whether each replica can independently decide to terminate a transaction or if the replicas decide on the basis of a vote. Whereas the comparison in~\cite{wiesmann:2000:SRDS} is purely theoretical, our present work takes an experimental approach to comparison of popular databases. Sousa et al. implemented a tool to test the dependability and performance of group communication based database replication protocols~\cite{sousa:2005:DSN}. Our current work focuses on profiling specific popular database management systems instead of the underlying protocols.

Pati{\~n}o-Martinez et al. proposed a middleware assisted database replication strategy in~\cite{patino:2005:TOCS} and implemented it on PostgreSQL. Mann et al. proposed a Software Defined Networking(SDN) approach to service replication in a data center~\cite{mann:2013:IM}. They propose that instead of the applications managing their replication, the network takes this responsibility. Xu et al. proposed a deduplication and compression based approach to reducing network bandwidth requirements for a replicated document database~\cite{xu:2015:SCC}. In~\cite{zhuang:2015:SPEC}, Zhuang et al. studied the LinkedIn Databus protocol for database replication with the objective of ensuring low latency replication. Their findings indicate that it is possible to reasonably forecast future LinkedIn traffic. They also proposed that some headroom must be kept in capacity planning in order to ensure low latency replication.

Bandwidth requirement analysis for database replication is also performed in~\cite{boru:2015:ClusterComputing}, but it is based on mathematical models and simulation. The authors of~\cite{boru:2015:ClusterComputing} also proposed an algorithm for energy efficient replication. This algorithm monitors the access and update rates to database objects, estimates the future access and update rates and monitors congestion in a data center network to prioritize replicas to be updated in a given time interval.

In~\cite{minhas:2013:VLDB}, Minhas et al. proposed RemusDB, a technique for building highly available databases using the Remus high availability solution. RemusDB can work with any underlying database management system and provides low fail-over latency. They also perform a transaction processing performance analysis, which is not our focus in the present work.
\section{Experimental Setup}
\label{sec:setup}

\begin{figure}
\centering
\includegraphics[width=0.4\textwidth]{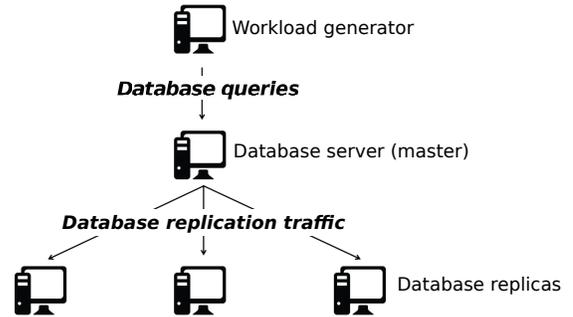}
\caption{Experimental setup.}
\label{fig:setup}
\end{figure}

The setup of our test-bed is shown in Fig.~\ref{fig:setup}. It consists of a workload generator alongwith a master database server and upto three database server replicas\footnote{Note that we do not count the master server in the number of replicas. When we say that there are two replicas, it means there are a total of three DBMS servers.}, also known as slaves. All of these were separate physical machines - Dell Precision T3500 workstations - connected over a Fast Ethernet LAN. The configuration of the machines in each of these roles is shown in TABLE~\ref{tab:setup}.

\begin{table}
\centering
\begin{tabular}{|c|c|c|c|}
\hline Role & IP Address & Hard disk & RAM (GB)\\
\ & \ & drive (GB) & \ \\
\hline Frontend & 192.168.50.11 & 500 & 18\\
\hline Database & 192.168.50.12 & 160 & 16\\
Server & \ & \ & \ \\
(master) & \ & \ & \ \\
\hline Database & 192.168.50.13 & 500 & 10\\
slave 1 & \ & \ & \ \\
\hline Database & 192.168.50.14 & 160 & 8\\
slave 2 & \ & \ & \ \\
\hline Database & 192.168.50.15 & 160 & 8\\
slave 3 & \ & \ & \ \\
\hline
\end{tabular}
\caption{PC configuration for the test-bed}
\label{tab:setup}
\end{table}

A program running on the workload generator was used to send commands to insert new entries into a database on the master server. The slaves were configured to be on standby, i.e., they were synced with the master but did not receive or respond to client queries directly. As changes were made on the master server's database, the same were replicated to the slave servers. Meanwhile, we captured network traffic, using tcpdump, on all the computers. We now describe the components of our experimental setup individually in more detail.

\subsection{Database Server (Master)}
\label{subsec:server} We ran separate sets of experiments for three popular DBMSs, namely, MySQL, PostgreSQL and Cassandra. A variety of third party tools for replication exist. However, in this study, we only used the replication built into these products. For experiments using MySQL, we used MySQL version $5.5.4$. For experiments using PostgreSQL, we used PostgreSQL version $9.3.16$. For experiments using Cassandra, we used version $2.2.9$. Each of these DBMS servers was installed on Ubuntu OS $14.04$ running on a Dell Precision T3500 workstation configured with $16$ GB RAM and a $160$ GB hard disk drive - the database master server machine (TABLE~\ref{tab:setup}). The database table schema varies with the data-set used in the experiment.

\subsection{Database Replicas (Slave)}
\label{subsec:slave} For each DBMS, we ran experiments while varying the number of replicas from $0$ to $3$. During an experiment, each of these slaves were running the same version of DBMS as that of corresponding master. The slaves were installed on Ubuntu 	$14.04$ OS running on three separate Dell Precision T3500 workstations with a minimum of $8$GB RAM and $160$GB HDD.

\subsection{Workload Generator}
\label{subsec:wlgenerator}
We used a separate system to generate the workload for the server. This was a Dell Precision T3500 workstation with $18$ GB RAM and $500$GB HDD. The workload generator runs a program that repeatedly inserts records into a database on the database master server. The nature and number of records inserted into the database varies with the data-set used. 

\subsection{Data-sets}
\label{subsec:dataset} We used three different datasets in our experiments. The first dataset consisted of a total of $10609$ tweets and re-tweets by a specific user captured between $29/12/2013$ and $18/01/2017$~\footnote{\label{datasets}The data-set and all test code is available at https://github.com/msaqib/DBRO/}. The average length of a tweet was $87$ characters. The total size of data-set on disk is approximately $30.6$ MB. That makes an average record size of $296.5$ bytes. Each record has $10$ columns. Accordingly, a database was created for inserting records from this data-set that has one table with $10$ columns including time-stamp, tweet ID, tweet text, client information as well as information about the tweet which is replied to or re-tweeted. This data-set was extracted using Twitter's web based interface for downloading one's own tweet archive.

The second dataset consisted of crime reports for Sacramento, CA, USA for the month of January, 2006~\cite{CrimesDataset}. This data-set consists of $7585$ records with $9$ fields including time-stamp, latitude and longitude of incident, address and crime description. This data-set was different from tweets since it did not contain long readable text, instead it consisted of short coded strings such as ‘6C’ and ‘123BURKLAY’. To insert records from this data-set, a database was created that has one table with $9$ columns, one for each field. The size of the data-set on disk was $793.6$ kB, which makes an average record size of $107.14$ bytes.

We also considered a dataset consisting of $25000$ flickr images~\cite{huiskes:2008:MIR}, since images are treated differently as compared to text by DBMSs. These images were inserted into a single table database. The average size of an image in this data-set was $117$ kB.


\subsection{Experiment structure}
\label{subsec:wlgeneration} We coded Python programs to run experiments that update the database on the database master server \ref{datasets} using our data-sets. The master database server was configured to replicate on up to $3$ slaves. For each combination of DBMS, data-set and a specific number of replicas, we ran an experiment at least $13$ times for statistical soundness. Each experiment consisted of the following activities: 
\begin{enumerate}
\item Insertion of records, one at a time, from the data-set into the database master server.
\item Replication of inserted records from the database master server to the replica(s). 
\item Execution of scripts that monitor /proc/stat, run diagnostic tools such as tcpdump and Gnome System Monitor to capture relevant network and resource usage footprints. 
\end{enumerate}
\section{Empirical Results}
\label{sec:results}Using the results of the experiments described in the previous section, we now compare the three DBMSs against various criteria. 

\subsection{Network Overhead}
\label{subsec:networkoverhead}

\begin{table}
\centering
\begin{tabular}{|c|c|c|c|}
\hline \ & \multicolumn{3}{|c|}{Total traffic between workload generator} \\
DBMS & \multicolumn{3}{|c|}{and database master server (MB)} \\
\cline{2-4} \ & Images data-set & Crimes data-set & Tweets data-set \\
\hline MySQL & 64.95 & 4.42 & \textbf{8.09} \\
\hline PostgreSQL & 125.92 & \textbf{2.8} & 10.51 \\
\hline Cassandra & \textbf{62.77} & 4.96 & 8.34 \\
\hline
\end{tabular}
\caption{Traffic between workload generator and database master server}
\label{tab:baseline}
\end{table}

For a given DBMS, we ran several experiments for each data-set. The average number of bytes exchanged between the workload generator and the database master server was calculated. These results are tabulated in In TABLE~\ref{tab:baseline}. For each data-set, the minimum number of bytes over the three DBMSs is highlighted in TABLE~\ref{tab:baseline}. For instance, for the images data-set, the minimum number of bytes exchanged between the workload generator and the database master was $62.77$ MB, for Cassandra. For a given data-set, we used the minimum number of bytes exchanged between the workload generator and database master server as the baseline when comparing the DBMSs in terms of total traffic on the network. Let $B_{Baseline}^z$ represent the baseline traffic for data-set $z$. Also, for data-set $z$, with DBMS $x$ configured with $y$ replicas, let $B_{x,y}^z$ represent the total traffic exchanged between the workload generator, the database master server and all $y$ slaves. We define the network overhead for DBMS $x$ with $y$ replicas for data-set $z$ as:
\begin{align}
O_{x,y}^z = {\frac{B_{x, y}^z - B_{Baseline}^z}{B_{Baseline}^z}}\times{100}, 
\label{eq:networkoverhead}
\end{align} 

For the images data-set, PostgreSQL had a much higher traffic between the workload generator and the database master than the other two DBMSs. In fact, the traffic between the workload generator and master server was about twice the physical size of the images on disk, which was surprising. On the other hand, for the same DBMS, the bytes being replicated from the master to a replica was of the same order as the size of the images on disk. Thus, something was off between the workload generator and PostgreSQL master server. We investigated this further and found the source of the problem in the PostgreSQL Python API that our workload generator script used to encode the images into SQL queries. Our script reads the binary image file one byte at a time. The value read from the disk was then being split into two bytes as it was encoded as a textual SQL string. Suppose the value of a byte was read as F4, then the script would insert the F and the 4 as separate characters in the SQL query string. Since a character occupies one byte, this which results in doubling in the size of data as it is transmitted from the workload generator to the database master server. We could not find any fix for this in the PostgreSQL Python API. However, re-writing the workload generator script using the PostgreSQL C API fixed this problem.
\begin{figure}
    \centering
    \begin{subfigure}[]{0.43\textwidth}
        \includegraphics[width=\linewidth]{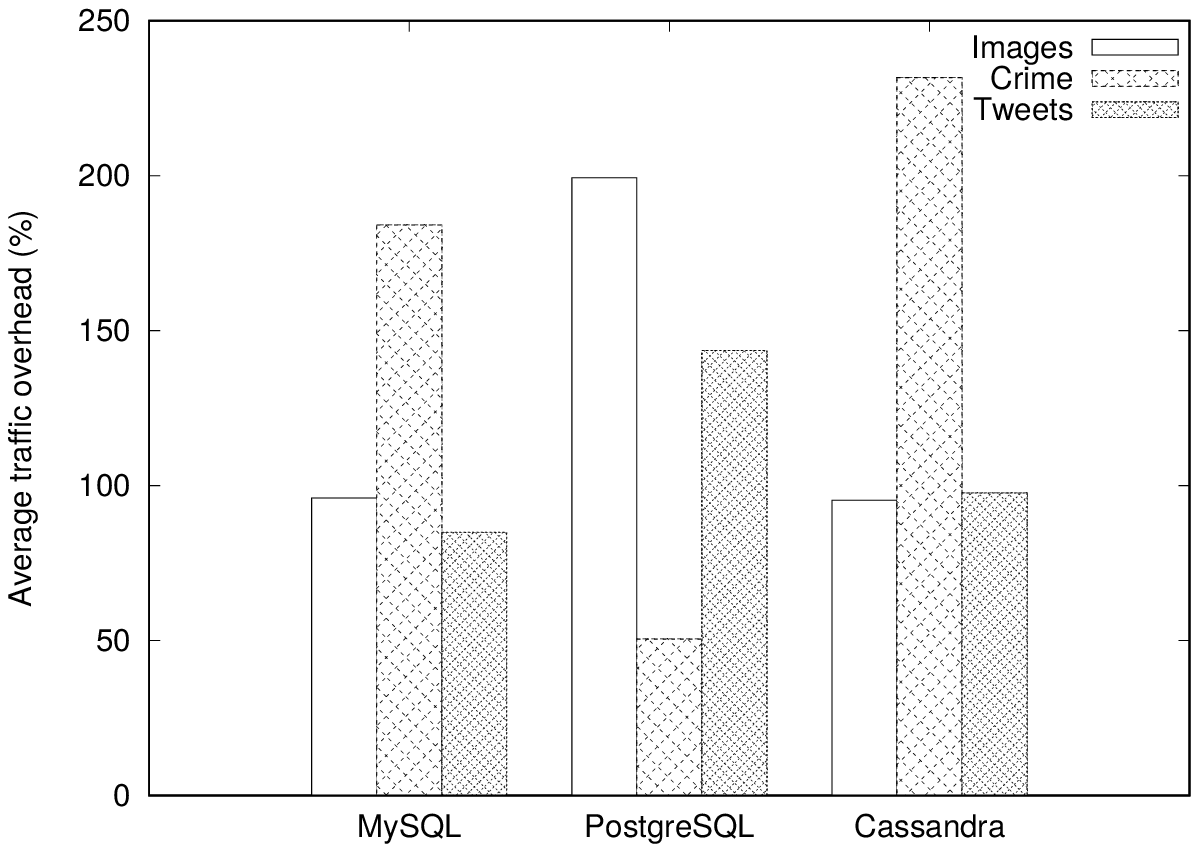}
\caption{One replica}
\label{fig:oneslave}
    \end{subfigure}
    ~ 
    \begin{subfigure}[]{0.43\textwidth}
        \includegraphics[width=\linewidth]{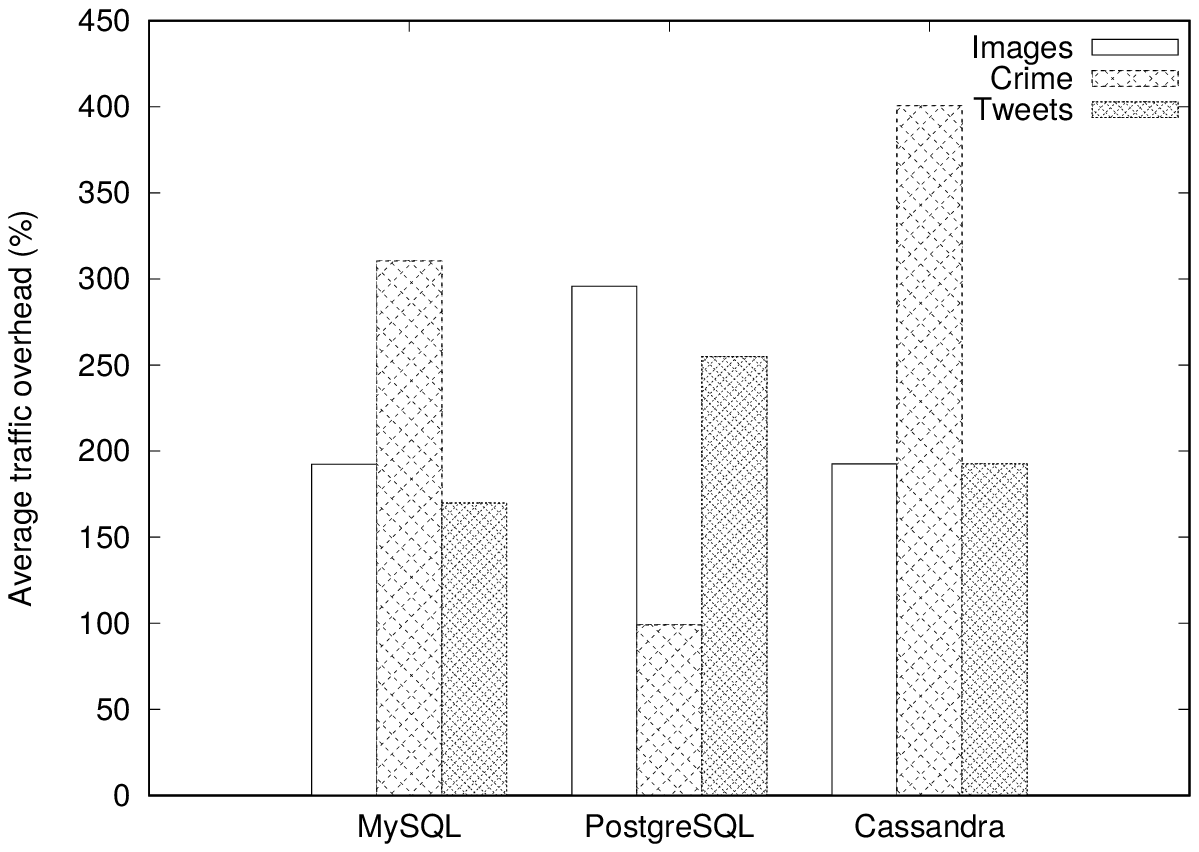}
\caption{Two replicas}
\label{fig:twoslaves}
    \end{subfigure}
    ~ 
    \begin{subfigure}[]{0.43\textwidth}
\includegraphics[width=\linewidth]{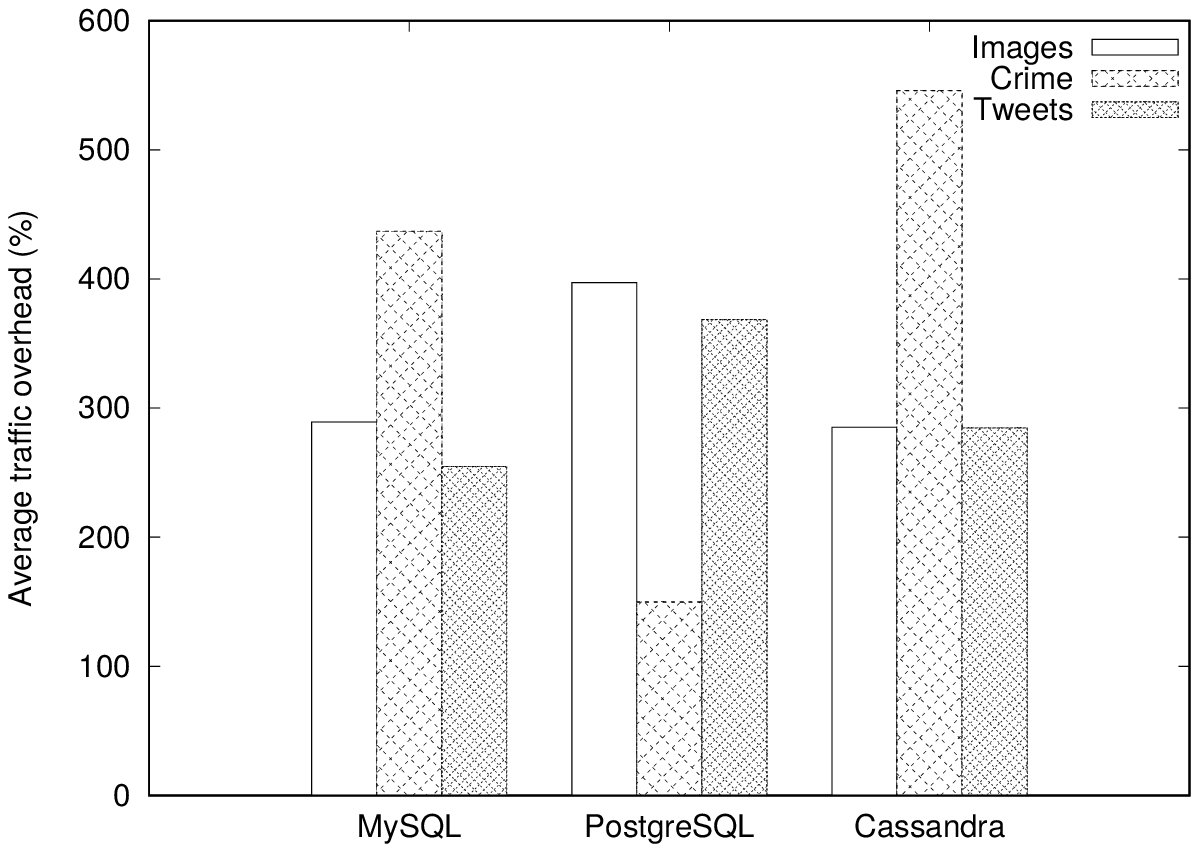}
\caption{Three replicas}
\label{fig:threeslaves}
    \end{subfigure}
    \caption{Percentage traffic overhead for varying number of replicas.}\label{fig:generic}
\end{figure}

For the crimes data-set, the traffic between workload generator and database master was lowest for PostgreSQL. On the other hand, for the tweets data-set, the traffic between workload generator and database master was the highest for PostgreSQL. It is surprising that PostgreSQL does well for one textual data-set and not the other. Upon investigation, we determined two factors at play here. First, the average record size for the crime data-set is smaller. Each crime record is more likely to fit in one encrypted data packet compared to a tweet record. Opening the packet traces in Wireshark shows packets labeled "\textbf{Application Data}" that carry the encrypted records from workload generator to master server and from master to the replicas. The number of application data packets in the packet trace going from the workload generator to the master server for the crime data-set was $7603$, which is nearly equal to the number of records in the crime data-set, i.e., $7585$. However, against the $10609$ records in the tweets data-set, a total of $31879$ application data packets were found going from the workload generator the master server. Thus, an average tweet fits in a greater number of application data packets compared to an average crime record. Each additional packet brings its own encryption and packetization overhead. The second factor resulting in a greater number of bytes on the network for the tweets data-set is that the experiment takes longer to complete for this data-set than the crime data-set. After every $5$ seconds or so, there are four packets that perform a new key exchange totaling about $2300$ bytes. So, there is some regular encryption related overhead as well.

Fig.~\ref{fig:generic} shows the percentage traffic overhead over the baseline traffic when the database has one, two and three replicas. Fig.~\ref{fig:oneslave},~\ref{fig:twoslaves} and~\ref{fig:threeslaves} show that, for the images data-set, the total traffic on the network is lowest for Cassandra, followed closely by MySQL, irrespective of the number of replicas. PostgreSQL has a very high traffic overhead for this data-set irrespective of the number of replicas. However, the main contributor to the high traffic overhead is the inefficient image insert operation implemented using the PostgreSQL Python API, as described earlier. If the traffic between the workload generator and the PostgreSQL database master were roughly halved for this data-set, as it should be, then the traffic overhead is nearly equal to that of MySQL.

For the crime data-set, irrespective of the number of replicas, PostgreSQL had the lowest overhead among the three DBMSs followed by MySQL and Cassandra. However, for the tweets data-set, PostgreSQL had the highest traffic overhead, for any number of replicas whereas MySQL had the lowest traffic overhead for the tweets data-set. 

Overall, the traffic overhead grows almost linearly with the number of replicas for all three DBMSs. Table~\ref{tab:trafficgrowth} shows the total traffic with two and three replicas as a fraction of the total traffic with one replica. For all data-sets, the total traffic on the network for MySQL with two slaves was around 1.4 times the total traffic on the network with one slave. When another MySQL replica was added, the total traffic became 1.9 times the traffic with one slave. The total traffic with two and three replicas as a fraction of the traffic with one replica grew in a somewhat similar fashion for PostgreSQL and Cassandra as well. The total traffic between the workload generator and the database master server always appears to be slightly greater than that between the master server and any of its replicas.

For the images data-set, the traffic overhead is higher for PostgreSQL than for MySQL for any number of replicas, as can be seen in Figures~\ref{fig:oneslave},~\ref{fig:twoslaves} and~\ref{fig:threeslaves}. Once again, the difference between the two DBMSs would not be so pronounced, if it weren't for the inefficiency in the Python API for PostgreSQL for this data-set. For the crime data-set, the PostgreSQL data-set had lower overhead than MySQL for any number of replicas. Meanwhile, for the tweets data-set, the overhead was lower for MySQL than PostgreSQL irrespective of the number of replicas. 

\begin{table}
\centering
\begin{tabular}{|c|c|c|c|}
\hline DBMS & data-set & Two replicas & Three replicas\\
\hline \ & Images & 1.49 & 1.98 \\
 MySQL & Crime & 1.44 & 1.89 \\
\ & Tweets & 1.46 & 1.92 \\ 
\hline \ & Images & 1.32 & 1.66 \\
 PostgreSQL & Crime & 1.32 & 1.66 \\
\ & Tweets & 1.46 & 1.92 \\
\hline \ & Images & 1.49 & 1.97 \\
Cassandra & Crime & 1.51 & 1.95 \\
\ & Tweets & 1.48 & 1.95 \\ 
\hline
\end{tabular}
\caption{Growth in total network traffic with increase in number of replicas as a fraction of total traffic with one replica}
\label{tab:trafficgrowth}
\end{table}

In MySQL, there is an option to compress the replication traffic. When this option was turned on, for the images data-set with one replica, the total traffic between the database master and slave reduced by $3.1\%$. This low improvement is understandable as our data-set comprised JPEG images, which are inherently compressed. However, for crime and tweets data-sets, turning on compression resulted in a significant reduction in traffic by $20\%$ and $13\%$, respectively. 

In PostgreSQL version $9.3$ and Cassandra version $2.2.9$, there was no available built-in option for compression of replication traffic. It is reported that such a feature has been added to PostgreSQL starting with version $9.5$~\cite{PostgreSQLCompression}. 

\subsection{CPU Utilization}
\label{subsec:cpuutilization}

\begin{figure}
    \centering
    \begin{subfigure}[b]{0.43\textwidth}
        \includegraphics[width=\linewidth]{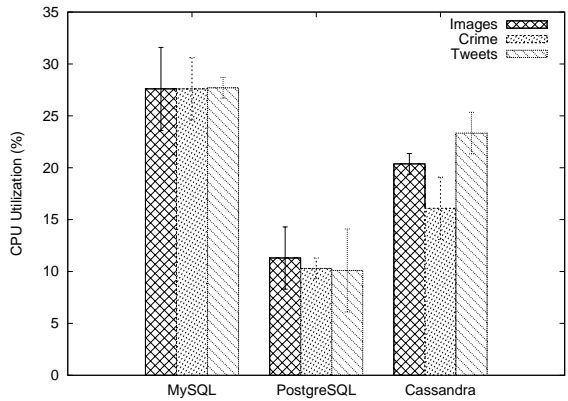}
\caption{One replica}
\label{fig:cpuutil1}
    \end{subfigure}
    ~ 
    \begin{subfigure}[b]{0.43\textwidth}
        \includegraphics[width=\linewidth]{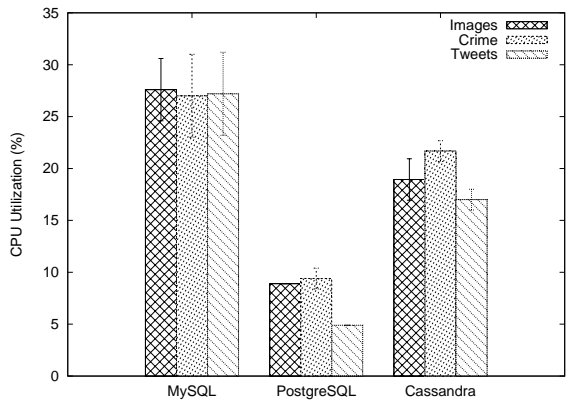}
\caption{Two replicas}
\label{fig:cpuutil2}
    \end{subfigure}
    ~ 
    \begin{subfigure}[b]{0.43\textwidth}
\includegraphics[width=\linewidth]{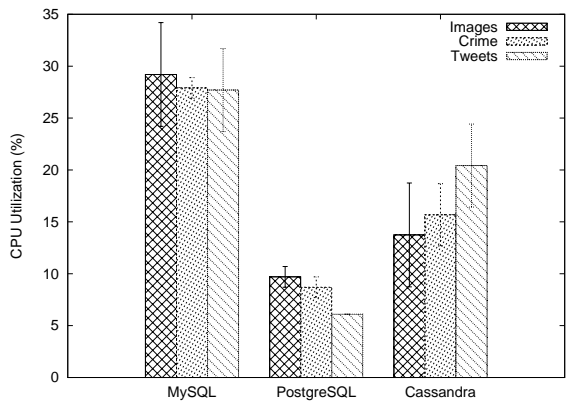}
\caption{Three replicas}
\label{fig:cpuutil3}
    \end{subfigure}
    \caption{CPU utilization for varying number of replicas with three different datsets.}\label{fig:cpuutil}
\end{figure}

Fig.~\ref{fig:cpuutil} shows the average, minimum and maximum CPU utilization recorded during several repetitions of the experiments using the three data-sets for all three DBMSs. It can be seen that no clear relationship exists between the number of replicas and the CPU utilization. However, the average CPU utilization for PostgreSQL is lowest amongst all three DBMSs irrespective of the data-set and number of slaves. Cassandra has the second highest average CPU utilization, whereas MySQL has the highest average CPU utilization. One implication of this result is that for our data-sets, a PostgreSQL backend would have the least electric energy requirement as power consumption is well approximated as an affine function of CPU utilization~\cite{Fan:ICSA:2007}.

When compressed replication was enabled in MySQL, we noticed that the average CPU utilization increased for all data-sets. The average CPU utilization for MySQL with compressed replication is reported for two of the data-sets\footnote{We were unable to collect the CPU utilization results for the tweets data-set with three replicas in time for this publication.} in TABLE~\ref{tab:cpuutilcompressedmysql}.

\begin{table}
\centering
\begin{tabular}{|c|c|c|c|}
\hline \ & \multicolumn{3}{|c|}{No. of slaves} \\
\cline{2-4} data-set & 1 & 2 & 3 \\
\hline Images & 29.11 & 33.33 & 32.4 \\
\hline Crime & 28.53 & 29.61 & 30.03 \\
\hline
\end{tabular}
\caption{Average CPU utilization during experiments with two different data-sets for MySQL with replication compression enabled.}
\label{tab:cpuutilcompressedmysql}
\end{table}

\subsection{Memory Utilization}
\label{subsec:memoryutilization}

The system's memory utilization during our experiments for all three DBMSs did not show a dependence on either the number of slaves or the data-set. In any given experiment, the system memory utilization was found to be nearly constant between $35\%$ and $40\%$.

\subsection{Time Consumption}
\label{subsec:timeconsumption}

\begin{figure}
    \centering
    \begin{subfigure}[b]{0.43\textwidth}
\includegraphics[width=\linewidth]{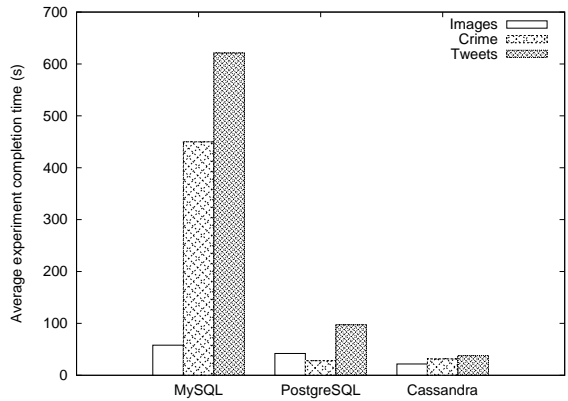}
\caption{One slave}
\label{fig:oneslavet}
    \end{subfigure}
    ~ 
    \begin{subfigure}[b]{0.43\textwidth}
\includegraphics[width=\linewidth]{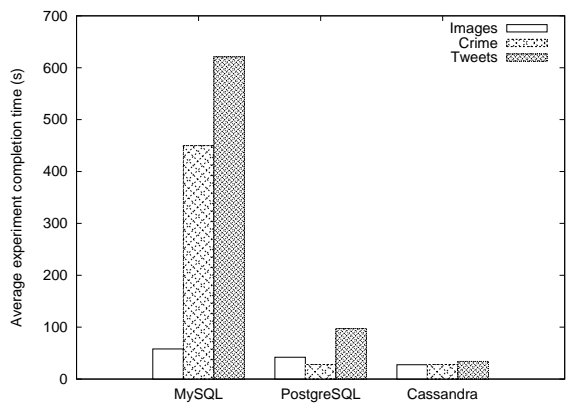}
\caption{Two slaves}
\label{fig:twoslavest}
    \end{subfigure}
    ~ 
    \begin{subfigure}[b]{0.43\textwidth}
\includegraphics[width=\linewidth]{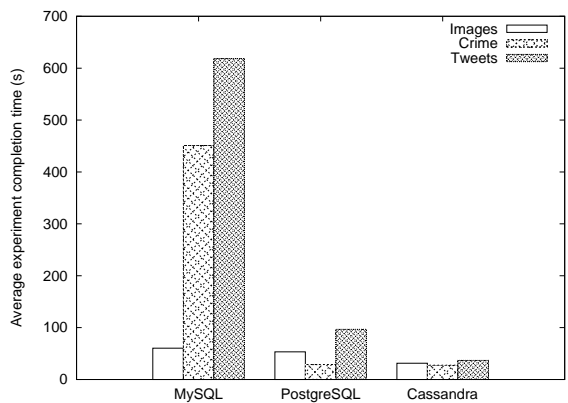}
\caption{Three slaves}
\label{fig:threeslavest}
    \end{subfigure}
    \caption{Average experiment completion time (s) with varying number of slaves.}\label{fig:generic2}
\end{figure}

We also measured the time required to insert the entire data-set into the database. It would be interesting to see if the number of replicas has any effect on this measure. Fig.~\ref{fig:oneslavet} shows the average experiment completion time for all three data-sets for the three DBMSs when one replica was configured. We see that MySQL is, in general, the slowest, but excessively slow for the textual data-sets. One possible reason for this sluggish performance could be TCP flow control limiting the throughput. Another possible reason could be greater server processing time, i.e., the time required by the server to process an insert operation before sending the reply packet to the workload generator. 

\begin{figure}
\centering
\includegraphics[width=0.45\textwidth]{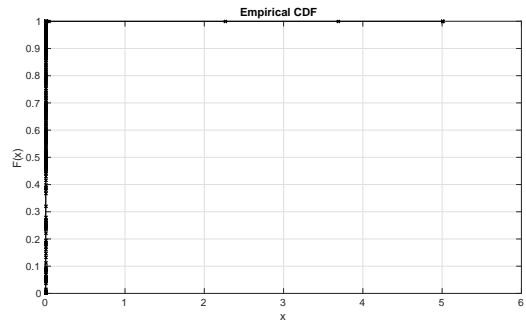}
\caption{Server processing time for PostgreSQL with the crime data-set.}
\label{fig:thinktimepostgrecrime}
\end{figure}

\begin{figure}
\centering
\includegraphics[width=0.45\textwidth]{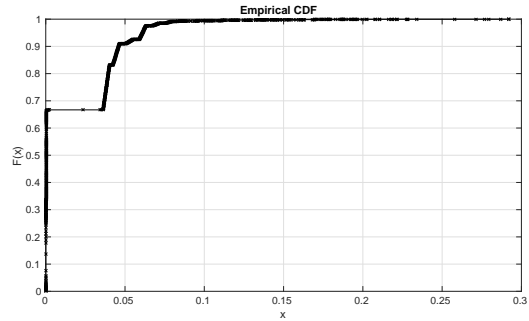}
\caption{Server processing time for MySQL with the crime data-set.}
\label{fig:thinktimemysqlcrime}
\end{figure}

We calculated the server processing time as the delay in the packet trace between the arrival of a request packet at the server and the corresponding response. For the crime data-set, the mean processing time for PostgreSQL was $1.6$ ms, whereas the same metric for MySQL for the same data-set was an order of magnitude higher at $16$ ms. The empirical CDF of the processing time for MySQL and PostgreSQL are plotted in Figure~\ref{fig:thinktimemysqlcrime} and~\ref{fig:thinktimepostgrecrime} respectively. It can be seen that for PostgreSQL nearly all requests incurred a very small processing time, with few packets incurring a large processing time. For MySQL, on the other hand, most packets had a significantly high processing time. The results for processing time reported here are from an experiment when one slave was considered for replication. Other results were also qualitatively similar.

Except for the crime data-set, Cassandra showed the best record insertion performance among the three DBMSs. Furthermore, comparing Figure~\ref{fig:oneslavet} with Figures~\ref{fig:twoslavest} and~\ref{fig:threeslavest} shows that increasing the number of slaves from $1$ to $3$ did not have a noticeable effect on the average experiment completion time for any of the DBMSs.

\section{Discussion}
\label{sec:discussion}
Based on our study, we find Cassandra to be most suitable for image intensive applications in terms of traffic overhead. For textual applications, PostgreSQL may be optimal depending on the average size of records being inserted. If the byte stream from the client-side is efficiently encoded, then PostgreSQL may be competitive for image oriented applications, too. PostgreSQL's traffic is encrypted by default, which is another positive. Also, the average CPU utilization of a server running PostgreSQL was found to be lower than the other two DBMSs, which is a positive with regard to energy consumption. 

In terms of rate of processing insertions into the database, Cassandra was found to be the best for all types of traffic. However, PostgreSQL's processing rate was quite close to that of Cassandra. If sheer performance is absolutely critical, then the choice appears to be Cassandra. However, if both processing rate and energy consumption are important, then PostgreSQL may be the DBMS of choice for any type of traffic. With the reported addition of compression of replication traffic in PostgreSQL version $9.5$, PostgreSQL's case could become stronger with reduced traffic overhead. However, further investigation is needed to evaluate that hypothesis. 
\section{Conclusions}
\label{sec:conclusions} We studied overheads of database replication, especially network traffic, for MySQL, PostgreSQL and Cassandra databases, using textual as well as image data-sets. We found that replication overhead traffic increases almost linearly with the number of replicas. We found that MySQL showed sluggish performance due to longer server think time. On the other hand, MySQL offers the ability to compress replication traffic. For our textual data-sets, this resulted in up to $20$\% reduction in total network traffic. An interesting finding of our study was the PostgreSQL's Python API is inefficient for storing binary images. Furthermore, PostgreSQL turned out to have the lowest average CPU utilization of all three DBMSs for all our data-sets, making it more energy efficient. 
\bibliographystyle{IEEEtran}
\bibliography{dbrep}

\end{document}